\documentclass{article}
\usepackage{url}
\usepackage{hyperref}
\usepackage{multicol}
\usepackage{datetime}
\usepackage{tikz}
\usepackage{mathtools}
\usepackage{subcaption}
\usepackage{enumerate}
\usepackage{todo}
\usepackage{microtype}%
\usetikzlibrary{patterns}
\usepackage[final]{pdfpages}

\usepackage{amsthm}
\usepackage{authblk}

\newtheorem{theorem}{Theorem}
\title{Flexible Paxos: Quorum intersection revisited}

\author[1,2]{Heidi Howard}
\author[1]{Dahlia Malkhi}
\author[1,3]{Alexander Spiegelman}
\affil[1]{VMware Research, Palo Alto, US\\
  \texttt{dahliamalkhi@gmail.com}}
\affil[2]{University of Cambridge Computing Laboratory, Cambridge, UK\\
  \texttt{heidi.howard@cl.cam.ac.uk}}
\affil[3]{Viterbi Dept. of Electrical Engineering, Technion
  Haifa, 32000, Israel\\
  \texttt{sashas@tx.technion.ac.il}}

\begin{document}

\maketitle

\begin{abstract}

Distributed consensus is integral to modern distributed systems. The widely adopted Paxos algorithm uses two phases, each requiring majority agreement, to reliably reach consensus. In this paper, we demonstrate that Paxos, which lies at the foundation of many production systems, is conservative. Specifically, we observe that each of the phases of Paxos may use non-intersecting quorums. Majority quorums are not necessary as intersection is required only across phases.

Using this weakening of the requirements made in the original formulation, we propose Flexible Paxos, which generalizes over the Paxos algorithm to provide flexible quorums. We show that Flexible Paxos is safe, efficient and easy to utilize in existing distributed systems. We conclude by discussing the wide reaching implications of this result. Examples include improved availability from reducing the size of second phase quorums by one when the number of acceptors is even and utilizing small disjoint phase-2 quorums to speed up the steady-state.

\end{abstract}

\section{Introduction}
\label{sec:intro}

Distributed consensus is the problem of reaching agreement in the face of failures. It is a common problem in modern distributed systems and its applications range from distributed locking and atomic broadcast to strongly consistent key value stores and state machine replication~\cite{schneider_cs90}. Lamport's Paxos algorithm~\cite{lamport_tcs98,lamport_sigact01} is one such solution to this problem and since its publication it has been widely built upon in teaching, research and practice.
At its core, Paxos uses two phases, each requires agreement from a subset of participants (known as a quorum) to proceed. The safety and liveness of Paxos is based on the guarantee that any two quorums will intersect. To satisfy this requirement, quorums are typically composed of any majority from a fixed set of participants, although other quorum schemes have been proposed.

In practice, we usually wish to reach agreement over a sequence of values, known as Multi-Paxos~\cite{lamport_sigact01}. We use the first phase of Paxos to establish one participant as a \emph{leader} and the second phase of Paxos to propose a series of values. To commit a value, the leader must always communicate with at least a quorum of participants and wait for them to accept the value.

In this paper, we weaken the requirement in the original protocol that all quorums intersect to require only that quorums from different phases intersect. Within each of the phases of Paxos, it is safe to use disjoint quorums and majority quorums are not necessary. We will refer to this new formulation as Flexible Paxos (FPaxos) as it allows developers the flexibility to choose quorums for the two phases, provided they meet the above requirement. FPaxos is strictly more general than Paxos and FPaxos with intersecting quorums is equivalent to Paxos.

Given that Multi-Paxos and its variants are widely deployed, such a result has
wide reaching practical applications. Since the second phase of Paxos
(replication) is far more common than the first phase (leader election), we can
use FPaxos to reduce the size of commonly used second phase quorums. For
example, in a system of 10 nodes, we can safely allow only 3 nodes to
participate in replication, provided that we require 8 nodes to participate when
recovering from leader failure. This strategy, decreasing phase 2 quorums at the
cost of increasing phase 1 quorums, is referred to in the body of the paper as \emph{simple
quorums}.

The simple quorum system reduces latency, as leaders will no longer be required to wait for
a majority of participants to accept proposals. Likewise, it improves steady
state throughput as disjoint sets of participants can now accept proposals,
enabling better utilization of participants and decreased network load.
The price we pay for this is reduced availability as the system can tolerate
fewer failures whilst recovering from leader failure.

Later, we will illustrate
that surprisingly, it is not always necessary to compromise availability for steady state performance. Examples include reducing the size of second phase quorums by one when the
number of acceptors is even, and utilizing quorum systems such as grid quorums, which decrease the quorum sizes of both phases.
In the following section we outline the basic Paxos algorithm using the standard terminology. Readers who are already familiar with the algorithm should proceed directly to the next section. In \S\ref{sec:fpaxos} we describe the observation in detail and then in \S\ref{sec:implications} motivate why such flexibility is useful in practice. \S\ref{sec:safety} gives an informal description of why it is safe to weaken Paxos's assumption on quorum intersection. In \S\ref{sec:eval} we evaluate a na\"{\i}ve implementation of FPaxos and demonstrate its usefulness. \S\ref{sec:reconfig} outlines how to dynamically choose quorums and \S\ref{sec:related} relates FPaxos to the existing work in the field. The appendix includes a TLA+~\cite{lamport_tla02} specification of the FPaxos algorithm which has been model checked against our safety assumption.

\section{Paxos}
\label{sec:paxos}

We wish to decide a single value $v$ between a set of processes. The system is asynchronous, each process may fail and the messages passed between them may be lost. Each process has one or more roles. We have three roles: the proposer, a process who wishes to have a particular value chosen, the acceptor, a process which agrees and persists decided values or the learner, a process wishing to learn the decided value.

A proposer who has a candidate value will try to propose the value to the acceptors. If a value has already been chosen, the proposer will instead learn it. The process of proposing a value has two stages: phase 1 and phase 2, each phase requires a majority of acceptors to agree in order to proceed. We will now look at each of these stages in details:

\noindent{\textbf{Phase 1 - Prepare \& Promise}}
\begin{enumerate}[i]
\item A proposer selects a unique proposal number p and sends \emph{prepare(p)} to the acceptors.
\item Each acceptor receives \emph{prepare(p)}. If p is the highest proposal number promised, then p is written to persistent storage and the acceptor replies with \emph{promise(p',v')}. \emph{(p',v')} is the last accepted proposal (if present) where p' is the proposal number and v' is the corresponding proposed value.
\item Once the proposer receives \emph{promise} from the majority of acceptors, it proceeds to phase two. Otherwise, it may try again with higher proposal number.
\end{enumerate}

\noindent{\textbf{Phase 2 - Propose \& Accept}}
\begin{enumerate}[i]
\item The proposer must now select a value v. If more than one proposal was returned in phase 1 then it must choose the value associated with the highest proposal number. If no proposals were returned, then the proposer can choose its own value for v. The proposer then sends \emph{propose(p,v)} to the acceptors.
\item Each acceptor receives a \emph{propose(p,v)}. If p is equal to or greater than the highest promised proposal number, then the promised proposal number and accepted proposal is written to persistent storage and the acceptor replies with \emph{accept(p)}.
\item Once the proposer receives \emph{accept(p)} from the majority of acceptors, it learns that the value v is decided.
Otherwise, it may try phase 1 again with a higher proposal number.
\end{enumerate}

Paxos guarantees that once a value is decided, the decision is final and no different value can be chosen. Paxos will reach agreement provided that $\left \lfloor{n/2}\right \rfloor +1 $ acceptors out of $n$ acceptors are up and are able to communicate. Proving progress requires us to make some assumptions about the synchrony of the system, as we cannot guarantee progress in a truly asynchronous systems~\cite{fischer_jacm85}.

Usually, we wish to reach agreement over a sequence of values, which we will refer to as slots. We could use distinct instances of Paxos to decide each value in the sequence i.e. the $i^{th}$ slot is decided by the $i^{th}$ instance of Paxos. In practice however, we can do much better and this is referred to as Multi-Paxos.

The first phase of Paxos is independent of the value proposed for any given instance, therefore phase 1 can be executed prior to knowledge of which value to propose. Furthermore, we can aggregate phase 1 over a series of slots. We refer to a proposer who has completed phase 1 as a \emph{leader}. To avoid loss of generality, we introduce another agent, the \emph{client} who is the origin of values for proposal. Clients may be external to the system or co-located with other processes such as the proposers.

\begin{figure}
  \centering
  \includegraphics[width=0.6\linewidth]{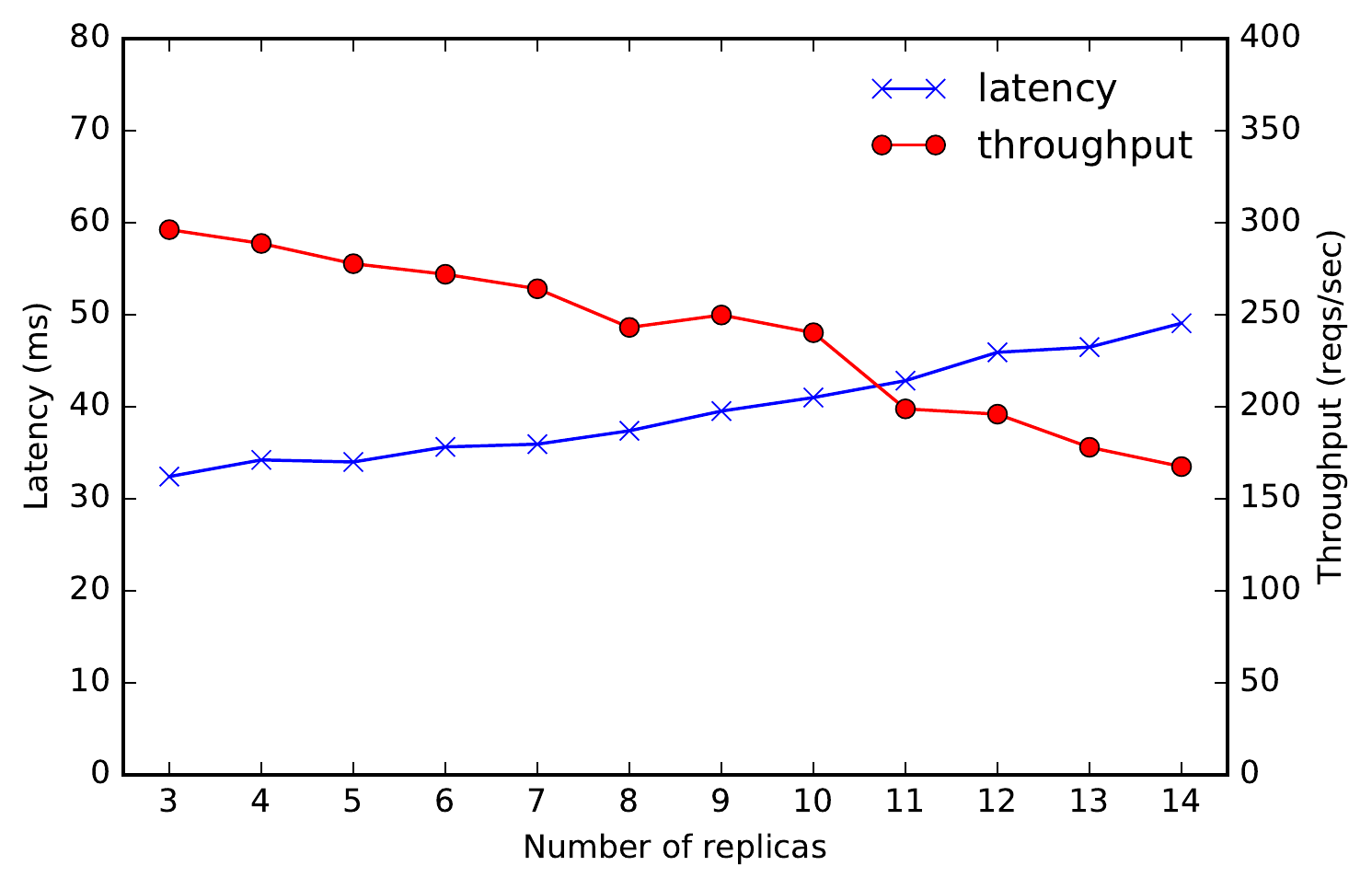}
  \caption{Performance of LibPaxos3 for varying system sizes. Details of the experimental setup are given in \S\ref{sec:eval}.}
  \label{fig:paxos}
\end{figure}

Figure~\ref{fig:paxos} illustrates how Multi-Paxos performs in practice. The x-axis shows the number of replicas in the system, each replica performs the roles of proposer, acceptor and learner. The blue line indicates the commit latency observed by the client and the red line indicates the average request throughput. As we would expect, increasing the number of replicas will increase latency and decrease throughput. These findings are consistent with previous studies~\cite{marandi_dsn10,mao_osdi08}.

A \emph{quorum system} is the method by which we choose which sets of acceptors are able to form valid quorums. It has been observed that Paxos can be generalized to replace majority quorums with any quorum system which guarantees that any two quorums will have a non-empty intersection~\cite{lamport_sigact01,lamport_msr05}. The fundamental theorem of quorum intersection states that its resilience is inversely proportional to the load on (hence the throughput of) participants~\cite{naor_siam98}. Therefore, with Paxos and its intersecting quorums, one can only hope to increase throughput by reducing the resilience, or vice versa. In the rest of this paper, we show that by weakening the quorum intersection requirement, we can break away from the inherent trade off between resilience and performance.

\section{FPaxos}
\label{sec:fpaxos}

In this section, we observe that the usual description of Paxos (as given in \S\ref{sec:paxos}) is more conservative than is necessary. To explain this observation, we will differentiate between the quorum used by the first phase of Paxos, which we will refer to as \emph{Q1} and the quorum for second phase, referred to as \emph{Q2}.

Paxos uses majority quorums of acceptors for both \emph{Q1} and \emph{Q2}. By requiring that quorums contain at least a majority of acceptors we can guarantee that there will be at least one acceptor in common between any two quorums. Paxos's proof of safety and progress is built upon this assumption that all quorums intersect.

We observe that it is only necessary for phase 1 quorums (\emph{Q1}) and phase 2 quorums (\emph{Q2}) to intersect. There is no need to require that \emph{Q1}'s intersect with each other nor \emph{Q2}'s intersect with each other. We refer to this as Flexible Paxos (FPaxos) and it generalizes the Paxos algorithm. If we allow any set of at least $\left \lfloor{n/2}\right \rfloor +1 $ acceptors to form a \emph{Q1} or \emph{Q2} quorum in FPaxos, then FPaxos is equivalent to Paxos.

Using this observation, we can make use of many non-intersecting quorum systems. In its most straight-forward application, we can simply decrease the size of \emph{Q2} at the cost of increasing the size of \emph{Q1} quorums.

As we discussed earlier, the second phase of Paxos (replication) is far more frequent than the first phase (leader election) in Multi-Paxos. Therefore, reducing the size of \emph{Q2} decreases latency in the common case by reducing the number of acceptors required to participate in replication, improves system tolerance to slow acceptors and allows us to use disjoint sets of acceptors for higher throughput. The price we pay for this is requiring more acceptors to participate when we need to establish a new leader. Whilst electing a new leader is a rare event in a stable system, if sufficient failures occur that we cannot form a \emph{Q1} quorum, then we cannot make progress until some of the acceptors recover.

Like Paxos, the system is able to make progress provided that at least enough acceptors are up and able to communicate to form both \emph{Q1} and \emph{Q2} quorums. Unlike Paxos, we are able to make progress within a given phase, provided we are able to form quorums corresponding to that phase. More concretely, if sufficient failures have occurred such that a proposer can no longer form \emph{Q1} quorums but is able to form the smaller \emph{Q2} quorums, the system can continue to safely make progress until a new leader is required. If the acceptors recover before the current leader fails, then the system suffers no loss in availability as a result.

\section{Implications}
\label{sec:implications}

We will now consider the practical implication of observing that quorums
intersection is required only between the two phases of Paxos. There already
exists an extensive literature on quorum systems from the fields of databases
and data replication, which can now be more efficiently applied to the field of
consensus. Interesting example systems include weighted
voting~\cite{gifford_sosp79}, hierarchies~\cite{kumar_trans91} and crumbling
walls~\cite{peleg_podc95}. For now however, we will illustrate the utility of
FPaxos by considering three na\"{\i}ve example quorum systems: (1) majority quorums; (2) simple quorums and (3) grid quorums.

\subsection{Majority quorums}
Currently, Paxos requires us to use quorums of size $n/2+1$ when the number of
acceptors $n$ is even\footnote{Lamport observed that majorities can be extended
to include exactly half of the sets of size $n/2$~\cite{lamport_nets76}.}. Using
our observation, we can safely reduce the size of \emph{Q2} by one from $n/2+1$
to $n/2$ and keep \emph{Q1} the same. Such a change would be trivial to
implement and by reducing the number of acceptors required to participate in replication, we can reduce latency and improve throughput. Furthermore, we have also improved the fault tolerance of the system. As with Paxos, if at most $n/2-1$ failures occur then we are guaranteed to be able to make progress. However unlike with Paxos, if exactly $n/2$ acceptors fail and the leader is still up then we are able to continue to make progress and suffer no loss of availability.

\begin{figure}
  \centering

  \begin{subfigure}[b]{0.4\textwidth}
    \centering
\begin{tikzpicture}
  \def\top{8}
  \def\bottom{0}

  \def\propA{0}
  \def\propAtext{\propA+1}
  \def\propB{5.5}
  \def\propBtext{\propB-1}

  \def\accA{2}
  \def\accB{2.5}
  \def\accC{3}
  \def\accD{3.5}

  \draw[thick] (\propA,\top) -- (\propA,\bottom) node[anchor=north] {P1};
  \draw (\accA,\top) -- (\accA,\bottom) node[anchor=north] {A1};
  \draw (\accB,\top) -- (\accB,\bottom) node[anchor=north] {A2};
  \draw (\accC,\top) -- (\accC,\bottom) node[anchor=north] {A3};
  \draw (\accD,\top) -- (\accD,\bottom) node[anchor=north] {A4};
  \draw[thick] (\propB,\top) -- (\propB,\bottom) node[anchor=north] {P2};

  \def\stepA{\top-.8}
  \node[above] at (\propAtext,\stepA) {prepare(1)};
  \draw[thick,->] (\propA,\stepA) -- (\accA,\stepA);
  \draw[thick,->] (\propA,\stepA-0.1) -- (\accB,\stepA-0.1);
  \draw[thick,->] (\propA,\stepA-0.2) -- (\accC,\stepA-0.2);

  \def\stepB{\stepA-.8}
  \node[above] at (\propAtext,\stepB) {promise()};
  \draw[thick,<-] (\propA,\stepB) -- (\accA,\stepB);
  \draw[thick,<-] (\propA,\stepB-0.1) -- (\accB,\stepB-0.1);
  \draw[thick,<-] (\propA,\stepB-0.2) -- (\accC,\stepB-0.2);

  \def\stepC{\stepB-1.2}
  \node[above] at (\propAtext,\stepC) {propose(1,a)};
  \draw[thick,->] (\propA,\stepC) -- (\accA,\stepC);
  \draw[thick,->] (\propA,\stepC-0.1) -- (\accB,\stepC-0.1);

  \def\stepD{\stepC-.7}
  \node[above] at (\propAtext,\stepD) {accept()};
  \draw[thick,<-] (\propA,\stepD) -- (\accA,\stepD);
  \draw[thick,<-] (\propA,\stepD-0.1) -- (\accB,\stepD-0.1);

  \def\stepE{\stepD-.8}
  \node[above] at (\propBtext,\stepE) {prepare(2)};
  \draw[thick,->] (\propB,\stepE) -- (\accD,\stepE);
  \draw[thick,->] (\propB,\stepE-0.1) -- (\accC,\stepE-0.1);
  \draw[thick,->] (\propB,\stepE-0.2) -- (\accB,\stepE-0.2);

  \def\stepF{\stepE-.8}
  \node[above] at (\propBtext,\stepF) {promise()};
  \draw[thick,<-] (\propB,\stepF) -- (\accD,\stepF);
  \draw[thick,<-] (\propB,\stepF-0.1) -- (\accC,\stepF-0.1);

  \def\stepG{\stepF-.7}
  \node[above] at (\propBtext,\stepG) {promise(1,a)};
  \draw[thick,<-] (\propB,\stepG) -- (\accB,\stepG);

  \def\stepH{\stepG-1.0}
  \node[above] at (\propBtext,\stepH) {propose(2,a)};
  \draw[thick,->] (\propB,\stepH) -- (\accD,\stepH);
  \draw[thick,->] (\propB,\stepH-0.1) -- (\accC,\stepH-0.1);

  \def\stepI{\stepH-.7}
  \node[above] at (\propBtext,\stepI) {accept()};
  \draw[thick,<-] (\propB,\stepI) -- (\accD,\stepI);
  \draw[thick,<-] (\propB,\stepI-0.1) -- (\accC,\stepI-0.1);

\end{tikzpicture}
     \caption{FPaxos with two serial proposals}
    \label{fig:fpaxos_example/1}
  \end{subfigure}

  \begin{subfigure}[b]{0.4\textwidth}
    \centering
\begin{tikzpicture}
  \def\top{6}
  \def\bottom{0}

  \def\propA{0}
  \def\propAtext{\propA+1}
  \def\propB{5.5}
  \def\propBtext{\propB-1}

  \def\accA{2}
  \def\accB{2.5}
  \def\accC{3}
  \def\accD{3.5}

  \draw[thick] (\propA,\top) -- (\propA,\bottom) node[anchor=north] {P1};
  \draw (\accA,\top) -- (\accA,\bottom) node[anchor=north] {A1};
  \draw (\accB,\top) -- (\accB,\bottom) node[anchor=north] {A2};
  \draw (\accC,\top) -- (\accC,\bottom) node[anchor=north] {A3};
  \draw (\accD,\top) -- (\accD,\bottom) node[anchor=north] {A4};
  \draw[thick] (\propB,\top) -- (\propB,\bottom) node[anchor=north] {P2};

  \def\stepA{\top-.8}
  \node[above] at (\propAtext,\stepA) {prepare(1)};
  \draw[thick,->] (\propA,\stepA) -- (\accA,\stepA);
  \draw[thick,->] (\propA,\stepA-0.1) -- (\accB,\stepA-0.1);
  \draw[thick,->] (\propA,\stepA-0.2) -- (\accC,\stepA-0.2);

  \def\stepB{\stepA-.8}
  \node[above] at (\propAtext,\stepB) {promise()};
  \draw[thick,<-] (\propA,\stepB) -- (\accA,\stepB);
  \draw[thick,<-] (\propA,\stepB-0.1) -- (\accB,\stepB-0.1);
  \draw[thick,<-] (\propA,\stepB-0.2) -- (\accC,\stepB-0.2);

  \def\stepC{\stepB-.8}
  \node[above] at (\propBtext,\stepC) {prepare(2)};
  \draw[thick,->] (\propB,\stepC) -- (\accD,\stepC);
  \draw[thick,->] (\propB,\stepC-0.1) -- (\accC,\stepC-0.1);
  \draw[thick,->] (\propB,\stepC-0.2) -- (\accB,\stepC-0.2);

  \def\stepD{\stepC-.8}
  \node[above] at (\propBtext,\stepD) {promise()};
  \draw[thick,<-] (\propB,\stepD) -- (\accD,\stepD);
  \draw[thick,<-] (\propB,\stepD-0.1) -- (\accC,\stepD-0.1);
  \draw[thick,<-] (\propB,\stepD-0.2) -- (\accB,\stepD-0.2);

  \def\stepE{\stepD-1.2}
  \node[above] at (\propAtext,\stepE) {propose(1,a)};
  \draw[thick,->] (\propA,\stepE) -- (\accA,\stepE);
  \draw[thick,->] (\propA,\stepE-0.1) -- (\accB,\stepE-0.1);

  \node[above] at (\propBtext,\stepE) {propose(2,b)};
  \draw[thick,->] (\propB,\stepE) -- (\accD,\stepE);
  \draw[thick,->] (\propB,\stepE-0.1) -- (\accC,\stepE-0.1);

  \def\stepF{\stepE-.7}
  \node[above] at (\propAtext,\stepF) {accept()};
  \draw[thick,<-] (\propA,\stepF) -- (\accA,\stepF);

  \node[above] at (\propBtext,\stepF) {accept()};
  \draw[thick,<-] (\propB,\stepF) -- (\accD,\stepF);
  \draw[thick,<-] (\propB,\stepF-0.1) -- (\accC,\stepF-0.1);

\end{tikzpicture}
     \caption{FPaxos with two concurrent proposals}
    \label{fig:fpaxos_example/2}
  \end{subfigure}

  \caption{Sample executions of FPaxos using improved majority quorums. The system is comprised of four acceptors (A1-A4) and two proposers (P1,P2)}
  \label{fig:fpaxos_example}
\end{figure}
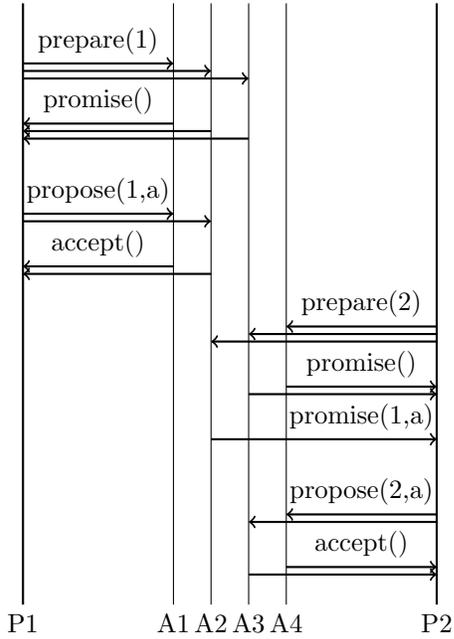
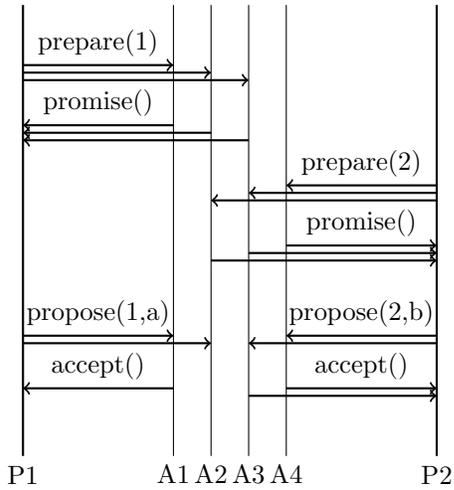

Figure~\ref{fig:fpaxos_example} shows two example traces of FPaxos with majority quorums in practice. As the system is comprised of four acceptors, FPaxos uses a majority (3 acceptors) for \emph{Q1} but requires only two acceptors for \emph{Q2}. In the examples, the two proposers wish to commit conflicting proposals. In figure~\ref{fig:fpaxos_example/1}, proposer one is first to execute FPaxos and its value \emph{a} is committed. Later, proposer two executes a round of Paxos and learns the value. In figure~\ref{fig:fpaxos_example/2}, both proposers successfully execute the first phase of FPaxos and simultaneously submit conflicting proposed values to the disjoint sets of acceptors. Both \emph{Q2}s will intersect with the two \emph{Q1}s, so only one of them will be successful. The unsuccessful proposer can retry with a higher proposal number and learn the chosen value.

\subsection{Simple quorums}

We will use the term \emph{simple quorums} to refer to a quorum systems where any acceptor is able to participate in a quorum and each acceptor's participation is counted equally. Simple quorums are a straightforward generalization of majority quorums. Paxos requires that all quorums intersect, and therefore, as we have previously discussed, each quorum must contain at least a strict majority of acceptors to meet this requirement.

In contrast, FPaxos requires only that quorums from different phases intersect. Therefore, FPaxos with simple quorums must require that $\left\vert{Q1}\right\vert + \left\vert{Q2}\right\vert > N$. We know that in practice the second phase is much more common than the first phase so we allow $\left\vert{Q2}\right\vert<N/2$ and increase the size of $Q1$ accordingly. For a given size of \emph{Q2} and number of acceptors N, then minimum size of our first phase quorum is $\left\vert{Q1}\right\vert = N-\left\vert{Q2}\right\vert+1$.
FPaxos will always be able to handle up to $\left\vert{Q2}\right\vert-1$ failures. However, if between $\left\vert{Q2}\right\vert$ to $N - \left\vert{Q2}\right\vert$ failures occur, we can continue replication until a new leader is required.

As has been previously observed~\cite{lamport_dsn04}, we do not need to send \emph{prepare} and \emph{propose} messages to all acceptors, only to at least $\left\vert{Q1}\right\vert$ or $\left\vert{Q2}\right\vert$ acceptors. If any of these acceptors do not reply, then the leader can send the messages to more acceptors. This reduces the number of messages from $4 \times N$ to $(2 \times \left\vert{Q1}\right\vert) + (2 \times \left\vert{Q2}\right\vert)$. This comes at the cost of increased latency, as the leader may not choose the fastest acceptors and must retransmit when
failures occur.

\begin{figure}
  \centering
  \begin{subfigure}[b]{0.4\linewidth}
    \centering
    \begin{tikzpicture}
      \def\s{0.8}
      \foreach \x / \y in {1/0, 1/1, 1/2, 0/3, 1/3, 2/3, 3/3, 4/3}
        \draw[pattern=north west lines, pattern color=green!100] (\s*\x,\s*\y) rectangle (\s*\x+\s,\s*\y+\s);
      \foreach \x / \y in {3/0, 0/1, 1/1, 2/1, 3/1, 4/1, 3/2, 3/3}
        \draw[pattern=north east lines, pattern color=blue!100] (\s*\x,\s*\y) rectangle (\s*\x+\s,\s*\y+\s);
      \draw[step=\s,black] (0,0) grid (5*\s,4*\s);
    \end{tikzpicture}
    \caption{Paxos}
    \label{fig:grid_quorum/paxos}
  \end{subfigure}
  \begin{subfigure}[b]{0.4\linewidth}
    \centering
    \begin{tikzpicture}
      \def\s{0.8}
      \foreach \x / \y in {1/0, 1/1, 1/2, 1/3}
        \draw[pattern=north west lines, pattern color=green!100] (\s*\x,\s*\y) rectangle (\s*\x+\s,\s*\y+\s);
      \foreach \x / \y in {0/1, 1/1, 2/1, 3/1, 4/1}
        \draw[pattern=north east lines, pattern color=blue!100]  (\s*\x,\s*\y) rectangle (\s*\x+\s,\s*\y+\s);
      \draw[step=\s,black] (0,0) grid (5*\s,4*\s);
    \end{tikzpicture}
    \caption{FPaxos}
    \label{fig:grid_quorum/fpaxos}
  \end{subfigure}

  \caption{Example of using a 5 by 4 grid to form quorums for a system of 20 acceptors}
  \label{fig:grid_quorum}
\end{figure}
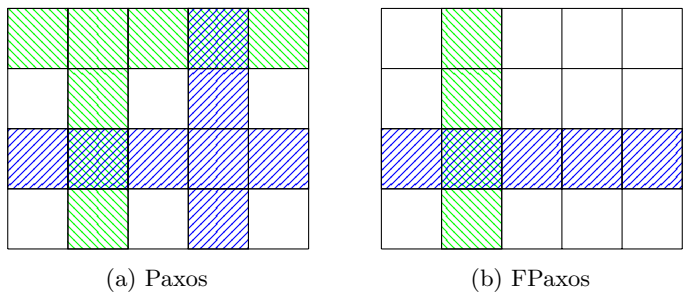

\subsection{Grid quorums}
The key limitation of simple quorums is that reducing the size of the \emph{Q2} requires a corresponding increase in the size of \emph{Q1} to continue to ensure intersection.
Grid quorums are an example of an alternative quorum system. Grid quorums can reduce the size of \emph{Q1} by offering a different trade off between quorum sizes, flexibility when choosing quorums and failure tolerance. Grid quorum schemes arrange the N nodes into a matrix of $N_{1}$ columns by $N_{2}$ rows, where $N_{1} \times N_{2} = N$ and quorums are composed of rows and columns. As with many other quorum systems, grid quorums restrict which combinations of acceptors can form valid quorums. This restriction allows us to reduce the size of quorums whilst still ensuring that they intersect.

Paxos requires that all quorums intersect thus one suitable grid scheme would require one row and one column to form a quorum\footnote{In practice, it is sufficient to use one row plus any choice of one grid item from each row below it. The average quorum size would become $N_{1}+(1/2)N_{2}$, although the worst case is still $N_{1}+N_{2}-1$.}. Figure~\ref{fig:grid_quorum/paxos} shows an example \emph{Q1} quorum and \emph{Q2} quorum using this scheme. This would reduce the size of a quorum from the majority of N to $N_{1}+N_{2}-1$. The number of failures which could be tolerated range from $MIN(N_{1},N_{2})$, where one node from every row or every column fails to $(N_{1}-1)\times(N_{2}-1)$, leaving only one row and one column remaining.

In FPaxos, we can safely reduce our quorums to one row of size $N_{1}$ for \emph{Q1} and one column of size $N_{2}$ for \emph{Q2}, examples are shown in Figure~\ref{fig:grid_quorum/fpaxos}. This construction is interesting as quorums from the same phase will never intersect, and may be useful in practice for evenly distributing the load of FPaxos across a group of acceptors.
With simple quorums, a system cannot recover from leader failure whilst any set
of $\left\vert{Q2}\right\vert = N/2$
acceptors have failed. Now with grid quorums, we are no longer treating all failures equally, it matters which of the acceptors have failed, not just how many have failed. Recall, that we are able to make progress in a given phase, provided we can still form a quorum for that phase.
For example, let us consider if four acceptors in either of grids from Figure~\ref{fig:grid_quorum} were to fail. If these failures occur across two columns then both systems will make progress. If all the failed nodes are within one column then no progress will be made by Paxos but FPaxos will continue until a new leader is needed. Likewise, if all the nodes in a given row where to fail, FPaxos would be able to complete \emph{Q1} and thus recover all past decisions, it can then safely fall back to a reconfiguration protocol to remove or replace the failed acceptors and continue to make progress.
In practice, failures are not independent and so we can distribute acceptors across the machines, racks or even data centers to minimize the likelihood of simultaneous failure.

By way of a thought experiment, let us consider setting $N_{1}=1$ and $N_{2}=N$ when using grid quorums or equivalently setting $\left\vert{Q1}\right\vert=N$ and $\left\vert{Q2}\right\vert=1$ with simple quorums. Any single acceptor will be sufficient to form a \emph{Q2}, however every acceptor must participate in \emph{Q1}. In practice, this would allow all acceptors to learn the decided value in a single hop, however we would be unable to recover from leader failure until every acceptor is up.

Alternatively, let us consider setting $N_{1}=N$ and $N_{2}=1$ when using grid quorums or equivalently setting $\left\vert{Q1}\right\vert=1$ and $\left\vert{Q2}\right\vert=N$ with simple quorums. This would require every acceptor to participate in \emph{Q2} but only a single acceptor is needed for \emph{Q1}. If any acceptors are still up, then we can complete \emph{Q1} and learn past decisions. As it has been previously observed~\cite{lamport_dsn04,lamport_podc09}, such a construction allows us to tolerate $f$ failures with only $f+1$ acceptors instead of $2f+1$.

\section{Safety}
\label{sec:safety}

Lamport's proof of safety for Paxos does not use the full strength of the
assumptions made, namely that all quorums will intersect. For the sake of completeness, in this section we outline the proof of safety for FPaxos.

For FPaxos to be safe, every decision that is reached must be final. In other words, once a value has been decided, no different value can be decided. This can be formally expressed as the following requirement:

\begin{theorem}
If value $v$ is decided with proposal number $p$ and $v'$ is decided with proposal number $p'$ then $v=v'$
\label{theorem1}
\end{theorem}

For a given value $v$ to be decided, it must first have been be proposed. Thus the following requirement is strictly stronger:

\begin{theorem}
If value $v$ is decided with proposal number $p$ then for any message \emph{propose(p',v')} where $p'>p$ then $v=v'$
\label{theorem2}
\end{theorem}

Proof is by contradiction, that is, assume $v\neq v'$. We will consider the
smallest proposal number $p'>p$ for which such a message is sent.

Let $\mathcal{Q}_1$ and $\mathcal{Q}_2$ be the sets of all valid phase 1 and
phase 2 quorums respectively and $\mathcal{A}$ be the set of acceptors. Quorums
are valid provided that:

\begin{gather} \label{eq:subset1}
  \forall Q_1  \in \mathcal{Q}_1 : Q_1 \subseteq \mathcal{A} \\
  \forall Q_2  \in \mathcal{Q}_2 : Q_2 \subseteq \mathcal{A}\\
  \forall Q_1 \in \mathcal{Q}_1, \forall Q_2 \in \mathcal{Q}_2 : Q_1 \cap Q_2 \neq \emptyset
\end{gather}

Equation 1 specifies that every possible phase 1 quorum is a subset of the
acceptors, likewise for equation 2. Equation 3 specifies that all possible
combinations consisting of a phase 1 and a phase 2 quorum will intersect in at least one acceptor.

Let ${Q}_{p,2}$ be the phase 2 quorum used by proposal number $p$ and
${Q}_{p',1}$ be the phase 1 quorum used by proposal number $p'$. Let $\bar{A}$
be the set of acceptors which participated both in the phase 2 quorum used by
proposal number $p$ and phase 1 quorum used by proposal number $p'$, thus
$\bar{A} = {Q}_{p,2} \cap {Q}_{p',1}$. Since ${Q}_{p,2} \in \mathcal{Q}_2$ and
${Q}_{p',1} \in \mathcal{Q}_1$ then we can use equation 3 to infer
that at least one acceptor must participate in both quorums, $\bar{A} \neq \emptyset$.

Let us consider the ordering of events from the perspective of one acceptor
$acc$ where $acc \in \bar{A}$. It is either the case that they receive \emph{prepare(p')} first or \emph{propose(p,v)} first.
We will consider each of these cases separately:

\textbf{CASE 1:}

Acceptor $acc$ receives \emph{prepare(p')} before it receives
\emph{propose(p,v)}. When $acc$ receives \emph{propose(p,v)}, its last promised
proposal will be $p'$ or higher. As $p'>p$ then it will not accept the proposal
from $p$, however as $acc \in {Q}_{p,2}$ it must accept \emph{propose(p,v)}. This is a contradiction thus it cannot be the case.

\textbf{CASE 2:}

Acceptor $acc$ receives \emph{propose(p,v)} before it receives
\emph{prepare(p')}.  When $acc$ receives \emph{prepare(p')}, there are two
cases. Either:

\textbf{CASE 2a:} The last promised proposal by acceptor $acc$ is already higher than $p'$. Then it
will not accept the prepare from $p'$, however as $acc \in {Q}_{p',1}$ it must
accept \emph{prepare(p')}. This is a contradiction thus it cannot be the case.

\textbf{CASE 2b:} The last promised proposal by acceptor $acc$ is less than $p'$
then it will reply with
\emph{promise(q,v)} where $p\leq q<p'$. The value $v$ will be the same the one $acc$
accepted with $p$, under the minimility hypothesis on $p'$.

\medskip

$acc \in {Q}_{p',1}$ therefore $promise(q,v)$ will be at least one of
the responses received by the proposer of $p'$. If this is the only accepted
value returned, then its value $v$ will be chosen. Other proposals may also be
received for members of ${Q}_{p',1}$. Recall that $p<p'$. For each other
proposal $(q',v'')$ received, either:

\textbf{CASE (i) $q'<q$:} These proposal will be ignored as the proposer must choose the value associated with the highest proposal.

\textbf{CASE (ii) $p'<q'$:} This case cannot occur as an acceptor will only reply to $prepare(p')$ when last promised is $<p'$.

\textbf{CASE (iii) $p<q'<p'$ :} For an acceptor to have accepted $(q',v'')$ then it
must have first been proposed. This is impossible by the minimality assumption on $p'$.

Thus the value $v$ will be chosen, in contradiction to the assumption that $propose(p', v')$ was sent.

\medskip

We have provided a 2 page formal specification of the single-valued FPaxos protocol in TLA+~\cite{lamport_tla02}. We model checked this specification with disjoint quorums and the requirement~\ref{theorem2} was preserved. The FPaxos TLA+ specification is only a minor adaptation of the Paxos specification, given in \cite{lamport_tla02}.

\section{Prototype}
\label{sec:eval}

\begin{figure}
  \centering
  \begin{subfigure}[b]{0.45\textwidth}
    \includegraphics[width=\linewidth]{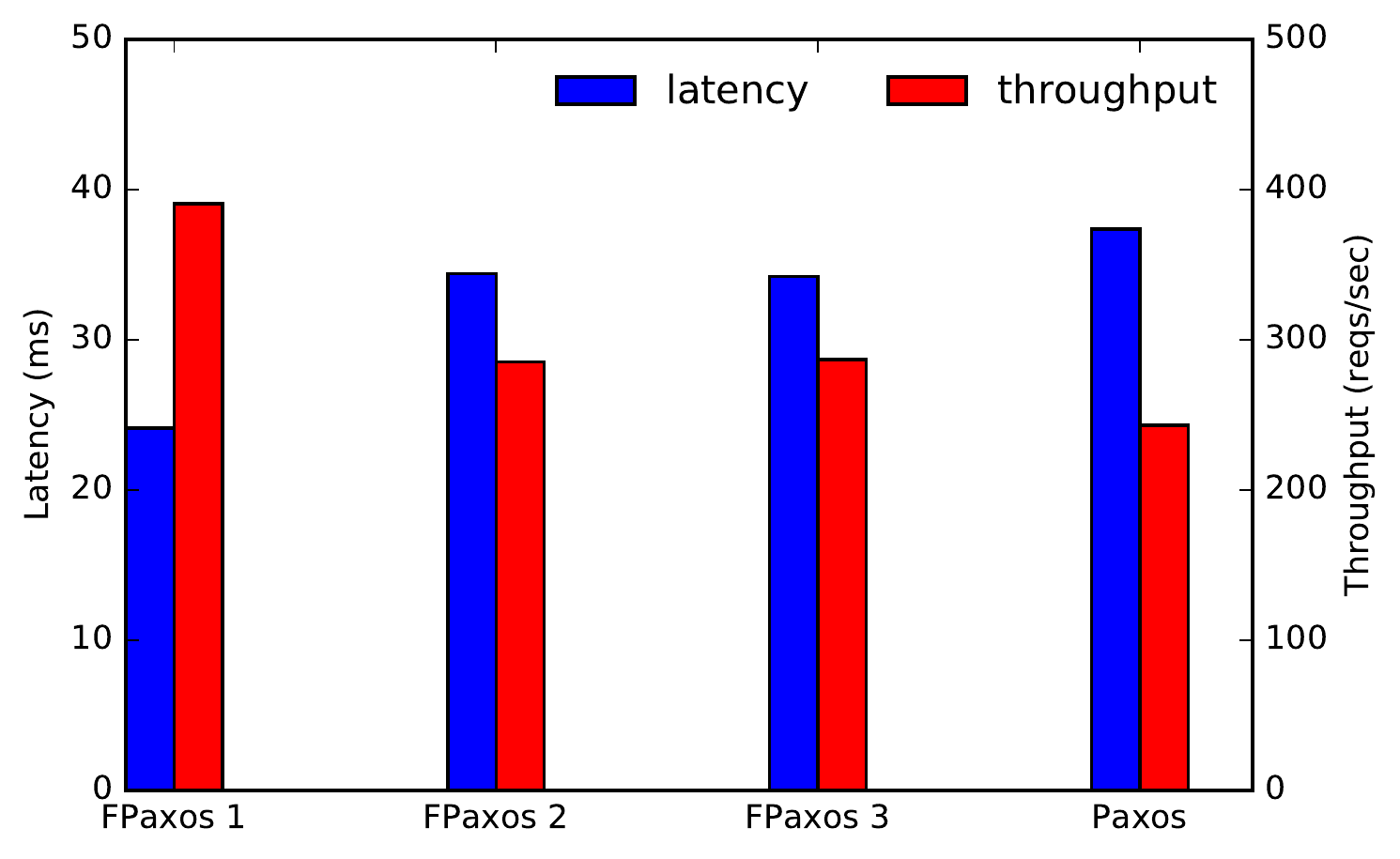}
    \caption{Performance of FPaxos and LibPaxos3 with 5 replicas.}
    \label{fig:fpaxos5}
  \end{subfigure}
\hspace{0.1in}
  \begin{subfigure}[b]{0.45\textwidth}
    \centering
    \includegraphics[width=\linewidth]{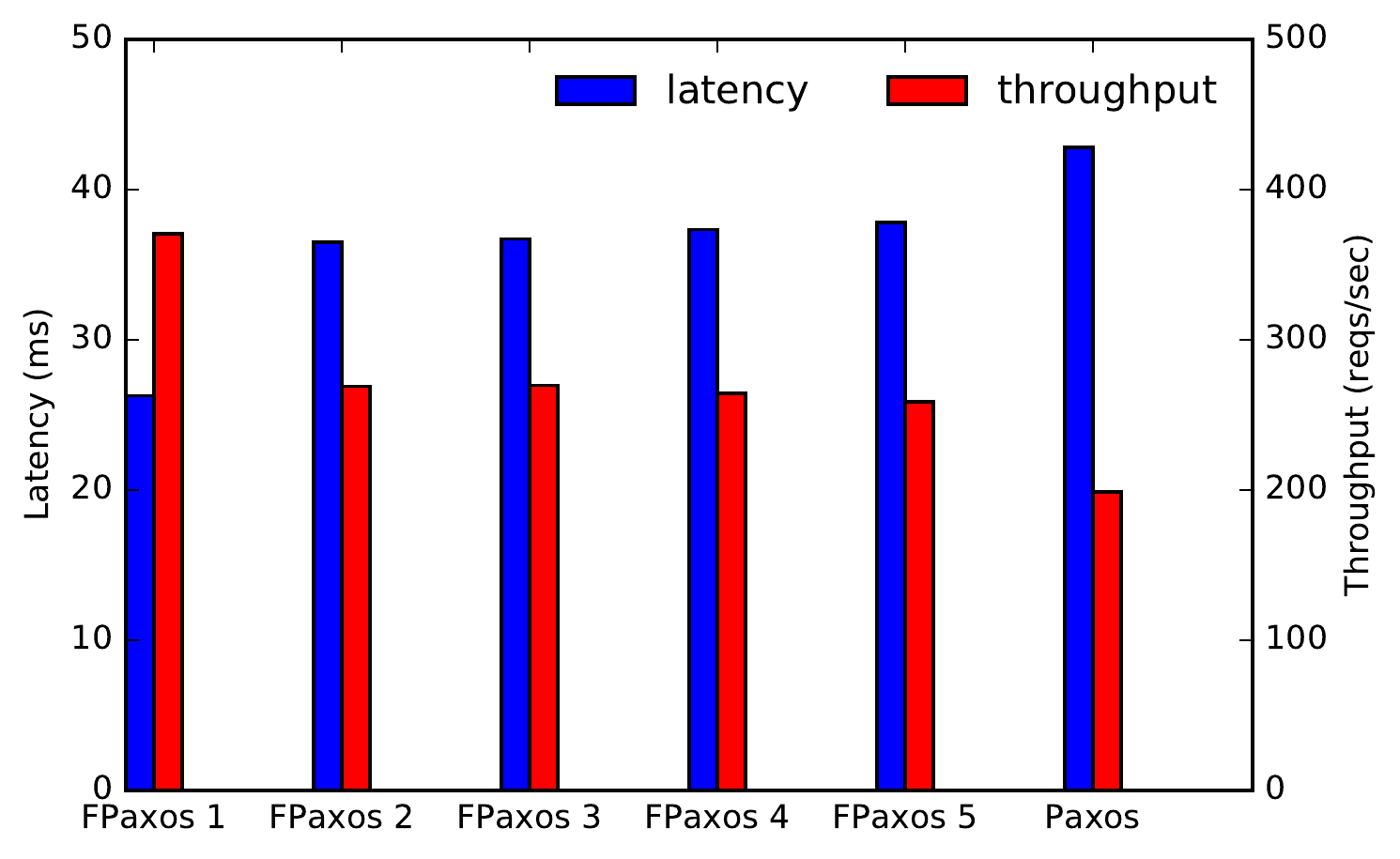}
    \caption{Performance of FPaxos and LibPaxos3 with 8 replicas.}
    \label{fig:fpaxos8}
  \end{subfigure}

  \caption{Throughput and average latency of FPaxos with various quorum sizes and LibPaxos3.}
  \label{fig:fpaxos_test}
\end{figure}

We implemented a na\"{\i}ve FPaxos by modifying LibPaxos3~\footnote{LibPaxos3 source code \url{https://bitbucket.org/sciascid/libpaxos}}, a commonly benchmarked Multi-Paxos implementation. Our modification simply generalized over the size of \emph{Q1} and \emph{Q2}. The simple quorums were na\"{\i}vely chosen at random and messages were sent only to a quorum of nodes.

LibPaxos3 is Multi-Paxos implementation in C which uses TCP/IP for transport. For each experiment, we tested N replicas, where each replica is a leader, an acceptor and a proposer. We used request sizes of 64 bytes with 10 requests in progress at any given time. Our experiments were ran within a single linux VM with a single core and 1GB of RAM, we used mininet~\footnote{\url{http://mininet.org}} to simulate a 10 Mbps network with 20 ms round trip time. Each test was run for 120 seconds, we discard the first and last 10 second to measure the system during its steady state.

Figure~\ref{fig:fpaxos_test} show the steady state performance of Paxos and FPaxos with varying \emph{Q2} quorum sizes. These results are as we would expect: by reducing the size of the \emph{Q2} quorum, we send fewer messages and thus increase throughput and decrease latency.

It is worth noting that this is not the complete picture.
First, FPaxos outperforms vanilla LibPaxos even with identical quorum sizes,
because FPaxos sends messages only to a quorum of replicas unlike LibPaxos3 which sends messages to all replicas.
When utilizing this optimization in practice, one may need to carefully trade the
strategy for finding quorums in realistic settings, and consider
replica failure, relative replica speeds and communication delays.
Second, unlike Paxos, FPaxos with \emph{Q2} of size 2 would not be able to elect a new leader when two acceptors have failed. On the other hand, in a system of 8 replicas, FPaxos with \emph{Q2} of size 4 handles more failures than Paxos, decreases latency (from 42ms to 37 ms) and increases throughput (from 198 to 264 reqs/sec).

This prototype demonstrates that implementing a na\"{\i}ve FPaxos is trivial. We show that even a very naive implementation improves performance and we believe that systems designed for FPaxos will see far greater performance, particularly by taking advantage of using disjoint set of acceptors and smarter quorum construction techniques to improve failure tolerance. Our prototype source code and associated materials are available online\footnote{\url{https://github.com/fpaxos}}.

\section{Enhancements}
\label{sec:reconfig}

We observe that the safety of FPaxos relies only on the assumption
that a given \emph{Q1} will intersect with all \emph{Q2}s with lower proposal numbers.
Therefore, we could further weaken the quorum requirements if a proposer was able to learn which \emph{Q2}s have been
used with smaller proposal numbers. We would then require only that a proposer's \emph{Q1} intersect
with these instead of all possible \emph{Q2}s.

In order to take advantage of this, we can enhance FPaxos with a mechanism for
leaders to select quorum(s) and to announce their selection. There are many ways
this could be implemented, but for safety the mechanism for a leader to make its quorum
selection known must be weaved carefully into the leader election protocol.
Details are left out of the scope of this paper. Briefly, it would be akin to Paxos reconfiguration and achieved by
adopting the principles of Vertical Paxos~\cite{lamport_podc09}.

The implications of this enhancement can be far reaching. For example, in a system of $N=100f$ nodes, a leader may start
by announcing a fixed \emph{Q2} of size $f+1$ and all higher proposal numbers
(and readers) will need to intersect with only this
\emph{Q2}. This allow us to tolerate $N-f$ failures.  Likewise, a leader may choose
a small set of $Q2$'s and announce all of them, allowing more flexibility in
phase 2 at the cost of less availability in phase 1. A
leader may also change its quorum selection over time using the dynamic
selection mechanism.

We expect that these enhancements and others may open many new possibilities for
practical system designs in the future.

\section{Related Works}
\label{sec:related}

The insightful State Machine Replication (SMR)
paradigm~\cite{lamport1978time, schneider1990implementing}
underlies many reliable systems, including pioneering works in
distributed systems field like Viewstamped
Replication~\cite{oki1988viewstamped} and
Isis~\cite{birman1987exploiting}.
The Paxos algorithm
provides the algorithmic solution for many production systems architected as
replicated state machines.
SMR must solve a core ingredient, agreement, which Dwork et
al.\cite{dwork1988consensus} solved under minimal synchrony
assumptions, and which is the basis for the single position
agreement protocol (called Synod) in
Paxos~\cite{lamport_tcs98}.
In the decades following its invention, the Paxos algorithm
has been extensively researched: it has been explained in
simpler terms~\cite{lamport_sigact01,van_cs15}, optimized for
practical systems~\cite{burrows_osdi06, hunt_atc10,
junqueira_dsn11, ongaro_atc14} and
extended to handle reconfiguration~\cite{lamport_podc09} and
arbitrary failures~\cite{castro_osdi99}.

Many variants of Paxos were proposed. Cheap
Paxos~\cite{lamport_dsn04} fixes a single phase 2 quorum until
a leader replacement occurs. Fast Paxos~\cite{lamport_msr05}
has a leaderless fast-path protocol which utilizes fast-track
phase 2 quorums of size $f + \lceil\frac{f+1}{2}\rceil$.
Mencius~\cite{mao_osdi08} uses a revolving leader regime.
Ring-Paxos~\cite{marandi2010ring, marandi2012multi} applies
the idea in Cheap Paxos~\cite{lamport_dsn04} to a ring overlay
using network-level multicast.
Chain Replication~\cite{van2004chain} daisy-chains acceptors
and collapses the two phases into one chain sweep. Generalized
Paxos~\cite{lamport_msr05b} extends state-machine replication
with commutative commands, and Egalitarian
Paxos~\cite{moraru_sosp13}
extends Generalized Paxos with fast-track quorums whose size
is $f+\lfloor\frac{f+1}{2}\rfloor$.
EVE~\cite{kapritsos2012all} optimistically concurrently agrees on
commands and later resolves conflicts in case they do not
commute.
Corfu~\cite{balakrishnan_tcs13} lets the
leader delegate its exclusive authority to any proposer in
order to yield better parallelism.
There are many other variants; a comprehensive taxonomy of
Paxos variants is given in~\cite{VanRenesse}.
These previous works were built on the foundations presented in
the pioneering protocols~\cite{oki1988viewstamped,
birman1987exploiting, dwork1988consensus, lamport_tcs98}, and focused on
enhancing them in order to achieve better performance.
Our new observation revisited the foundations and generalized
them; it is completely orthogonal and can be integrated into
previous protocols as well as to real production systems in order to
further improve performance.

The SMR reconfiguration problem was addressed in several
previous works.
Some use consensus commands to agree on next
configurations~\cite{oki1988viewstamped,
lamport2010reconfiguring, lamport_podc09}, whereas others use
the first phase to determine which quorum (out of a fix set of
quorums) will be used in the second phase~\cite{lamport_dsn04,
marandi2010ring}.
A general framework for reconfiguration that separates the
steady state agreement mechanism from the reconfiguration
event appears in~\cite{lamport_podc09}.
Reconfiguration for other fault tolerant services was also
previously investigated, e.g., in~\cite{gilbert2010rambo, aguilera_jacm11, jehl_isdc15, spiegelman_opodis15,
gafni2015elastic}.
As discussed in Section~\ref{sec:reconfig}, the ideas in these
works can be adopted in order to enhance FPaxos into a
reconfigurable and dynamic system.

To the best of our knowledge, we are the first to prove and implement this
generalization of Paxos. During the preparation of this publication,
Sougoumarane independently made the same observation on which this work is based
and released a blog post summarizing~\cite{sougoumarane} it for the systems
community.

\section{Conclusion}
\label{sec:conc}

In this paper we have described FPaxos, a generalization of the widely adopted Paxos algorithm, which no longer requires that quorums from the same Paxos phase intersect. We believe this result has wide ranging consequences.

Firstly, over the last two decades Multi-Paxos has been widely studied, deployed and extended. Generalizing existing systems to use FPaxos should be quite straightforward. Exposing replication (phase 2) quorum size to developers would allow them to choose their own trade off between failure tolerance and steady state latency.

Secondly, by no longer requiring replication quorums to intersect, we have removed an important limit on scalability. Through smart quorum construction and pragmatic system design, we believe a new breed of scalable, resilient and performant consensus algorithms is now possible.

\subparagraph*{Acknowledgements.}

We wish to thank the following people for their feedback: Jean Bacon, Jon Crowcroft, Stephen Dolan, Matthew Grosvenor, Anil Madhavapeddy, Sugu Sougoumarane and Igor Zabloctchi.

\bibliography{ref}

\appendix
\includepdf[pages=-]{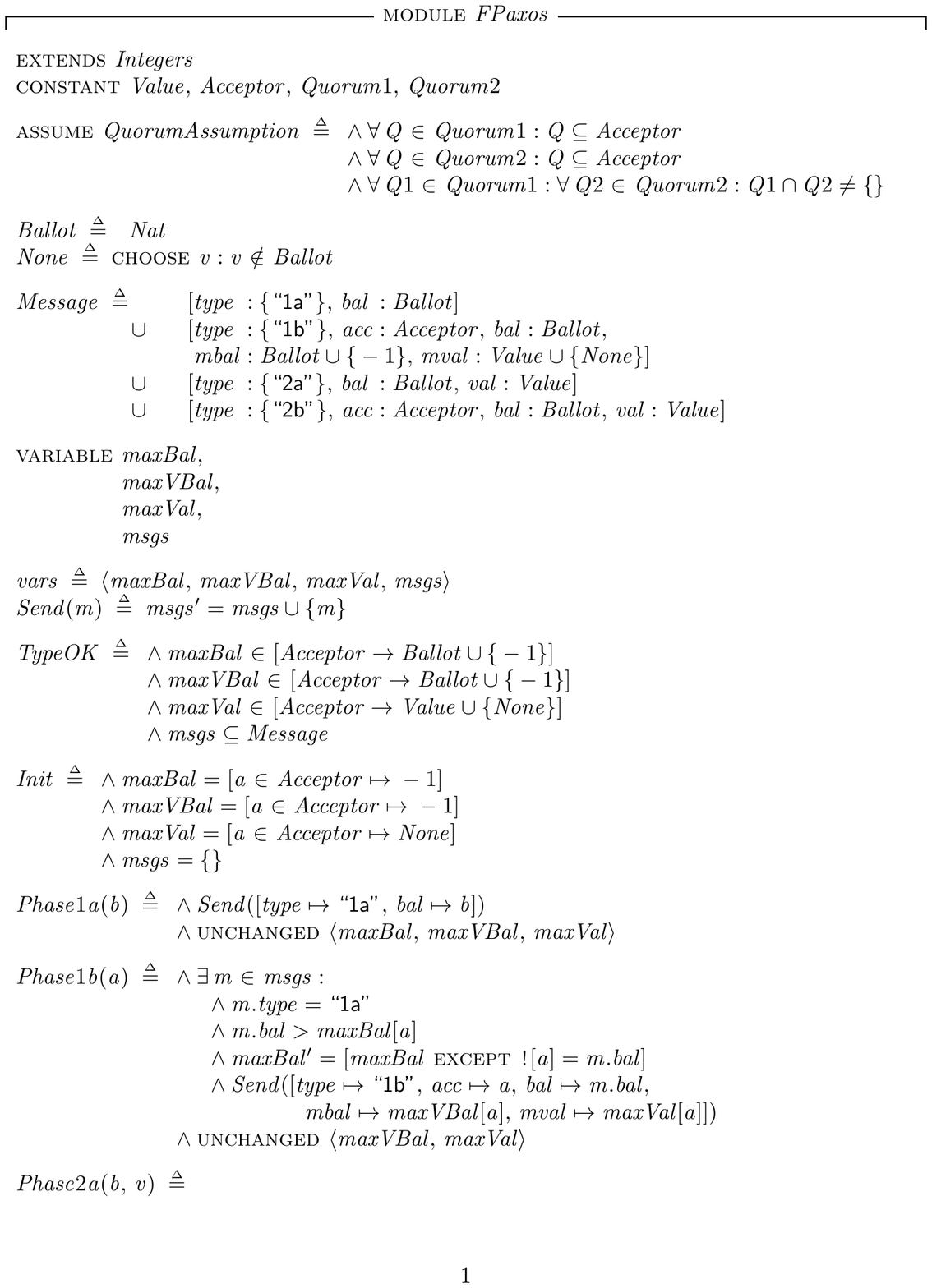}

\end{document}